\documentclass[12pt]{elsart}

\usepackage{graphicx}

\begin{document}

\begin{frontmatter}
\title{Determining the structure of Ru(0001) from low-energy electron diffraction of a single terrace}

\author[UAM]{J. de la Figuera,}
\ead[url]{http://hobbes.fmc.uam.es/loma}
\author[ICMM]{J. M. Puerta,}
\author[ICMM]{J. I. Cerda,}
\author[UAM]{F. El Gabaly,}
\author[Sandia]{and K. F. McCarty}

\address[UAM]{Dpto. de F\'{\i}sica de la Materia Condensada and Centro de Microan\'{a}lisis de Materiales, Universidad Aut\'{o}noma de Madrid, Madrid 28049, Spain}
\address[ICMM]{Instituto de Ciencia de Materiales de Madrid, CSIC, Madrid 28049, Spain}
\address[Sandia]{Sandia National Laboratories, Livermore, California 94551}
\date{\today}

\begin{abstract}

While a perfect hcp (0001) surface has three-fold symmetry, the diffraction patterns commonly obtained are six-fold symmetric. This apparent change in symmetry occurs because on a stepped surface, the atomic layers on adjacent terraces are rotated by 180 degrees.  Here we use a Low-Energy Electron Microscope to acquire the three-fold diffraction pattern from a single hcp Ru terrace and measure the intensity-vs-energy curves for several diffracted beams. By means of multiple scattering calculations fitted to the experimental data with a Pendry R-factor of 0.077, we find that the surface is contracted by 3.5$(\pm0.9)$\% at 456~K.
\end{abstract}

\begin{keyword}
Low energy electron diffraction (LEED)\sep
Low energy electron microscopy (LEEM)\sep
Ruthenium \sep
hydrogen\sep
Surface relaxation and reconstruction\sep
hexagonal close packed surface
\end{keyword}

\end{frontmatter}

\section{Introduction}
Low-energy electron diffraction (LEED) has been a workhorse technique for modern surface science\cite{vanhove}.  Experimental measurement of the intensity of the diffracted spots as a function of the electron's accelerating voltage (energy), I-V analysis, combined with full dynamical calculations of the electron scattering has become over the last two decades one of the most successful techniques for surface-structure determination. Although the development of synchrotron-radiation-related techniques, recent advances in scanning-probe microscopies, and the advent of efficient {\it ab-initio} formalisms have decreased its application, LEED remains a very powerful tool. The mature technique routinely determines the surface geometry with an accuracy well below a tenth of an Angstrom with error limits that are well understood\cite{tesis}.

Ruthenium has attracted interest for its catalytic properties\cite{RODRIGUEZ1991}. The clean hcp Ru(0001) surface does not reconstruct.  However, like most unreconstructed metal surfaces, the topmost layer of atoms relaxes inward towards the second layer. Both early LEED studies\cite{MICHALK1983} and more recent ones\cite{FEIBELMAN1994} reported that the first interplanar spacing, $d_1$, was contracted by 2\% relative to the bulk. However, {\it ab-initio} calculations by Feibelman\cite{FEIBELMAN1994,Feibelman1996} and Xu {\it et al.}\cite{Xu2005} found a much larger relaxation, between 3.5--4\%. This experiment/theory disagreement prompted suggestions that the smaller experimental relaxation may result from hydrogen adsorption\cite{FEIBELMAN1994}, although this possibility was disputed by others\cite{Menzel1994}. In a subsequent surface x-ray diffraction (SXRD) study, Baddorf {\it et al.}\cite{Baddorf2002} found a somewhat larger interplanar spacing and only small changes in $d_1$ upon H adsorption. Given this history, the Ru(0001) surface serves as a model system to test the accuracy of both experimental techniques and {\it ab-initio} calculations.

\begin{figure}
\centerline{\includegraphics[width=0.9\textwidth]{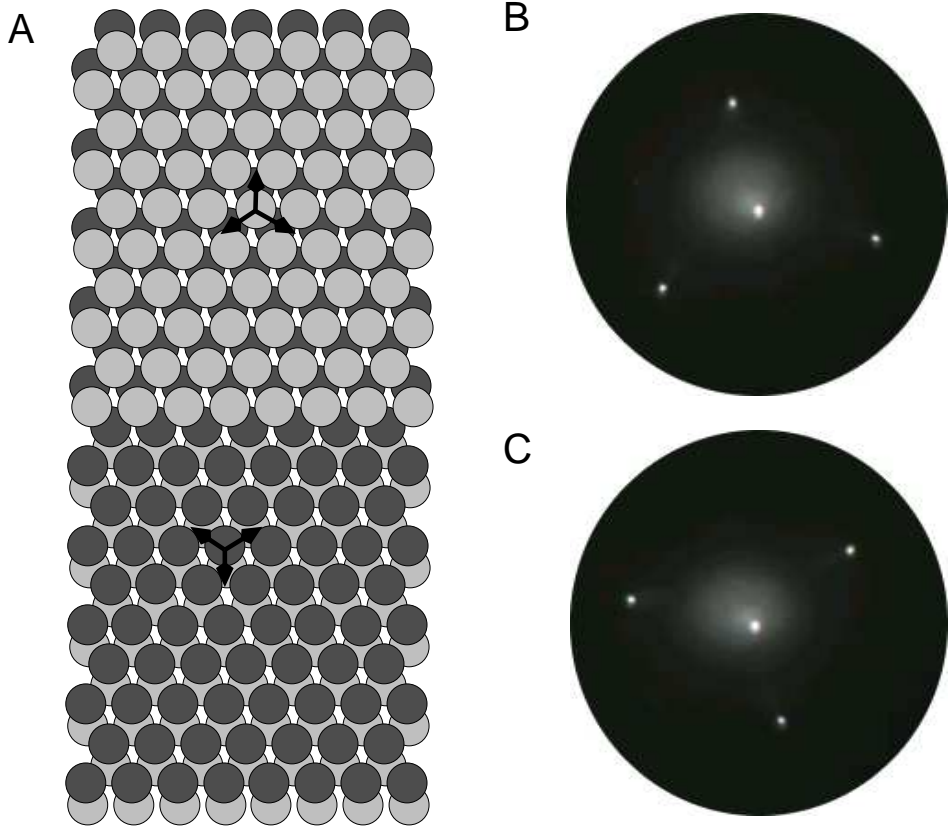}}
\caption{(a) Schematic illustration of two adjacent terraces on the Ru hcp surface separated by an atomic step. The arrows point from a surface atom to its three closest neighbors in the layer below.  Across adjacent terraces, the arrows are rotated by 180$^\circ$.
(b)-(c) LEED patterns at 32 eV acquired from two adjacent terraces showing
the three-fold symmetry of the surface and the rotation upon crossing 
a monoatomic step. The observed spots correspond to the specular and 
the integer (10) beams.}
\label{leed}
\end{figure}

As Fig.~\ref{leed} illustrates, the perfect (step-free) Ru(0001) surface exhibits three-fold ({\it p3m}) symmetry. Thus, diffraction from a single terrace should give a 3-fold pattern. However, most diffraction techniques show six-fold symmetry ({\it p6}) in the corresponding diffraction pattern\cite{Menzel1995,Pietro1998}. This apparent change in symmetry results from the different stacking on neighbor terraces, with termination ...BABA and ...ABAB respectively (where A and B indicate the two possible stacking positions of the hcp crystal). As Fig.~\ref{leed} also shows the surface terminations of adjacent terraces are rotated by 180$^\circ$ with respect to each other; therefore, diffraction from adjacent terraces yield 3-fold patterns that are rotated 180$^\circ$ with respect to each other. Thus, averaging over several terraces results in the commonly-observed six-fold symmetry.

In this work we revisit this surface to address the discrepancy between theory and experiment described above. We use the fine spatial resolution of low-energy electron microscopy (LEEM)\cite{Bauer1994,Bauer1998,Tromp2000} to obtain a new set of I-V curves from a single terrace on the Ru(0001) substrate.  In contrast, in traditional LEED measurements, the electron beam illuminates large areas comprising many terraces. Obviously, avoiding symmetry averaging should enhance the reliability of the I-V analysis. Indeed, our best-fit structure gives a first interplanar spacing, $d_1$, that is in excellent agreement with theoretical predictions within the carefully determined experimental errors.

\section{Experimental and Calculations Details}

The experiments were done in a commercial LEEM system with a base pressure of 7$\times 10^{-11}$ torr. The instrument has a lateral spatial resolution of 8 nm and uses a heated LaB$_6$ crystal as an electron source. The Ru(0001) substrate was cleaned by repeated cycles of oxygen adsorption and flash-annealing to 1700 $^\circ$C. The crystal contains a large area, about 200 $\mu$m wide, with a low density of atomic steps. In this region, terraces more than 5 $\mu$m wide can be found routinely. The video-rate acquisition system consists of a Peltier-cooled CCD camera whose output is digitized by a commercial digital video system at 30 frames of 720$\times$480 8-bit pixels per second. The Ru(0001) surface was aligned perpendicular to the electron beam using two independent tilt axes of a precision manipulator. The substrate/electron-beam alignment was further optimized by adjusting the microscope lenses so that the equivalent diffraction beams appeared/disappeared at exactly the same electron energy (i.e., the Ewald sphere was centered).

\begin{figure}
\centerline{\includegraphics[width=0.9\textwidth]{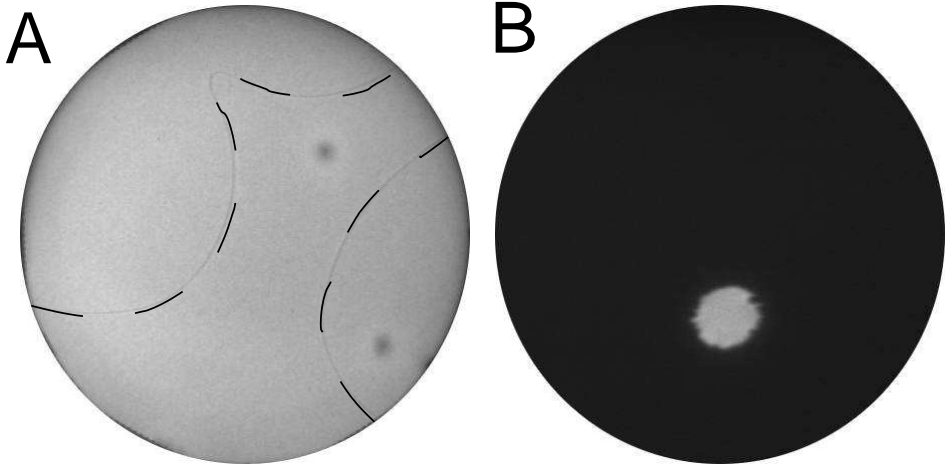}}
\caption{(a) LEEM image of the Ru surface. The field of view is 15~$\mu$m, and
the electron energy is 2.5~eV. The black dashed lines mark the two atomic steps present in the field of view (detected as faint gray lines). (b) LEEM image of the same area where a small aperture limits the electron beam to a 2.1~$\mu$m diameter region on a single terrace.  A diffraction pattern is formed from this selected area by changing the power of the microscope lenses.}
\label{leem}
\end{figure}

\begin{figure}
\centerline{\includegraphics[width=0.7\textwidth]{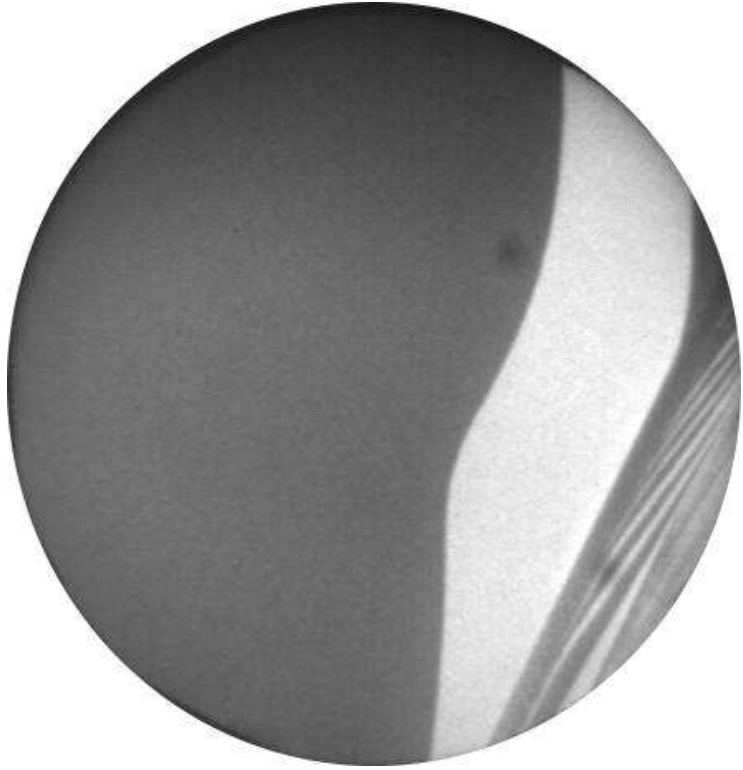}}
\caption{Dark-field image of the Ru(0001) surface. The field of view is 15~$\mu$m. The electron energy is 40~eV. An aperture blocks all the diffracted electrons except those corresponding to one of the first-order diffraction spots: only areas that
diffract electrons into the selected spot are imaged bright. Thus, the image intensity alternates between bright and dark when monoatomic steps are crossed.}
\label{dark}
\end{figure}

A major advantage of LEEM is the ability to image a surface and then obtain diffraction information from selected and well-characterized areas\cite{Bauer1994,Bauer1998,Tromp2000}.  This capability is illustrated in Fig.~\ref{leem}. Two curved monoatomic steps, which appear dark because of a phase-contrast mechanism\cite{Bauer1994,Chung1998}, are observed within the field of view of the left image. An area about two microns wide is selectively illuminated by inserting a smaller aperture into the illumination beam (right image). By changing the power of the imaging lenses, the diffraction pattern from this small area is obtained (Fig.~\ref{leed}).  Since this region contains only a single terrace (no steps), the LEED pattern is three-fold symmetric (Fig.~\ref{leed}b). The LEED pattern obtained from a region on an adjacent terrace is rotated by 180$^\circ$\ (Fig.~\ref{leed}c), consistent with the change in stacking between adjacent layers in an hcp crystal. Another way to show the symmetry change when crossing substrate steps is to use an aperture\cite{Bauer1994} to selectively form an image from a non-specular diffraction spot\cite{Farid2005}. In this dark-field imaging mode, only areas that diffract electrons into the selected spot are imaged bright in the real-space image. Thus, in an image (see Fig.~\ref{dark}) formed from an integer diffraction spot, adjacent terraces are alternately bright and dark. This dark-field contrast is similar to the one reported for w\"urtize (0001) surfaces\cite{Lagally2000}.

For diffraction measurements, a LEEM apparatus has several advantages over a conventional electron diffractometer. The electron path is well shielded from external magnetic fields, making very low-energy measurements possible (VLEED\cite{Lindroos1986}). Since there is no direct optical line of sight from sample to the detection screen, high-temperature measurements are straightforward. The LEEM geometry, where incoming and outgoing electrons are separated by a magnetic prism, allows measurement of the specularly reflected beam (i.e., the (00) beam) at normal incidence. The magnification of the LEED pattern can also be adjusted. Furthermore the spots do not move when the beam energy is changed. Unlike in conventional diffractometers, most current LEEM instruments do not perform energy filtering. Thus, secondary electrons are imaged together with the elastically scattered electrons. However, the secondary electrons can be easily removed by image processing because their contribution is spread out over a large area in reciprocal space, as described below.

The intensities of the (00), (01), (10) and (11) beams were obtained simultaneously by recording a sequence of images while changing the electron energy. The crystal was maintained at 456~K. To use the full dynamic range of the detection system and maximize the signal-to-noise ratio, the beam-energy range of 1 to 353~eV was split into three ranges (1-70~eV, 60-133~eV and 115-353~eV). Within each range, the electron current was adjusted to nearly saturate the image collection system at the most intense conditions. The different sections of the I-V curves were then rescaled with the same scaling factor for all the beams, by matching the intensity of overlapping scan regions. A plane was fit to the intensity of a box around the diffraction spot. This plane, which contains the secondary electron contribution, was then subtracted from the box. This procedure gives the same results as subtracting the average intensity along the box's perimeter, except where secondary electrons impose a large gradient on the background. The three (10) and (01) beams, as well as the six (11) beams, which are symmetry equivalent\cite{Lindgren}, where averaged together. Given that the (10) and (01) beams are not equivalent, we measured a total energy range of 1109~eV. The experimental I-V curves are shown in Fig.~\ref{iv}.

Full dynamical I-V calculations were performed with a modified version of the Van Hove-Tong package\cite{vanhove,tesis,huang,orh}. The surface was modeled by stacking two Ru atomic-planes on top of a Ru(0001) bulk. Well-converged values for both the number of beams and the number of phase shifts ($l_{max}=8$) were employed. In the structural search only normal displacements for the first two Ru atomic planes were considered ($d_1$ and $d_2$). The resulting parameter space was fully explored by calculating the I-V curves over a fine 2D grid where
$d_1$ and $d_2$ were swept over wide ranges. The experiment-theory agreement was quantified via Pendry's R-factor
\cite{pendry} ($R_P$), while the error bars on each parameter were obtained from its variance: $\Delta R_P = R_{P,min} \times \sqrt{ (8 V_i/\Delta E) }$, where $V_i$ gives the optical potential and $\Delta E$ corresponds to the total
energy range analyzed. Correlations between both parameters were taken into account for the error-limits estimation. Non-structural parameters such as the muffin-tin radius (r$_m$), the optical potential (V$_i$) and the Debye temperatures at the surface planes were also optimized ($\Theta_D^1$, $\Theta_D^2$ and $\Theta_D^{bulk}$ for the last layer, the next-to-last layer and the bulk, respectively), although they had no impact on the final surface geometry. The best-fit parameters are shown in Table~\ref{params}.

\section{Results and Discussion}

\begin{figure}
\centerline{\includegraphics[width=0.9\textwidth]{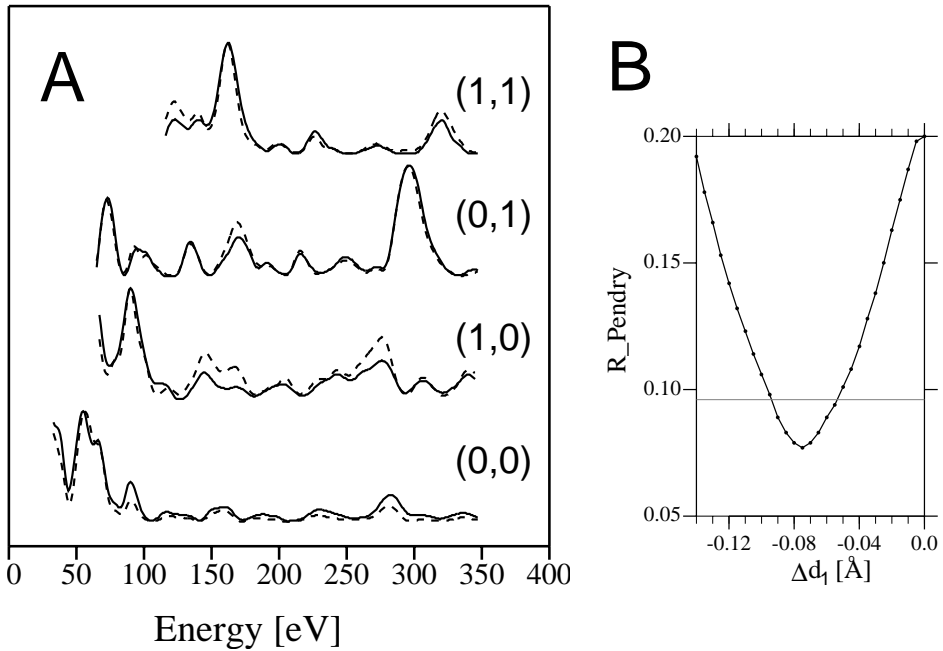}}
\caption{(a) Comparison between experimental (solid lines) and
calculated (dashed lines) I-V curves for the Ru(0001) surface. The total 
incident-beam energies spans a range of 1109~eV. (b) $R_P$ vs. first interlayer spacing, $d_1$. For each
fixed $d_1$ value in the graph, $R_P$ has been minimized as a function
of $d_2$, thus taking correlation effects into account. Gray horizontal
line gives the R-factor variance, $\Delta R_P$, from which the error
bars are estimated. }
\label{iv}
\end{figure}

The experimental I-V curves are shown in Fig.~\ref{iv}a. As is obvious from the data, the (01) and (10) beams are not equivalent. The best fit to the experimental data, shown on the same figure, was found for $d_1=2.065\pm 0.02$~\AA\  and $d_2=2.14\pm 0.025$~\AA.  With respect to the room temperature bulk value, $d_b=2.141$~\AA, these interplanar spacings give relaxations of $\delta d_1=-3.5(0.9)$\% and $\delta d_2=0.0(1)$\% (where the number in parenthesis indicates the error in percentage of the bulk value). The dependence of $R_P$ on $d_1$ is presented in Fig.~\ref{iv}b. The agreement between the experimental and simulated I-V curves is excellent, as reflected by an $R_P$ minimum value of $R_{P,min}=0.077$. The interplanar spacing of the topmost two layers for the optimized structure is in excellent agreement with the {\it ab-initio} calculations. That is, we find that $\delta d_1=-3.5(0.9)$\%  while theory finds 3.5-4\%\cite{FEIBELMAN1994,Xu2005}.  However, our structure differs somewhat from the previous experimental structures, which found $\delta d_1=-2$\%\cite{MICHALK1983,FEIBELMAN1994,Menzel1994} and $\delta d_1=-2.4(0.4)$\%\cite{Baddorf2002}.

Given the complexity of the LEEM optics, misalignment of the incidence angle, particularly changes with energy, would introduce an additional error in the derived structure.  We note that the intensities of the individual equivalent beams were well matched over the full energy range.  This observation strongly suggests that the incidence angle did not change with energy.  In addition, we explored the possibility of a systematic misalignment by optimizing the incidence angle in the R-factor analysis.  We found that $R_p$ had a pronounced minimum only at normal incidence and not for any off-normal condition.  The error bars of this analysis, even with the additional fitting parameter, was essentially the same as the best-fit structure of Fig.~\ref{iv} and Table~\ref{params}.  Thus, we conclude that the error in the best-fit structure due to the incidence angle is negligible.

We note that we report results at 456~K.  We investigated potential effects of this somewhat elevated temperature in two ways.  First, we acquired a complete, independent I-V data set at room temperature.  Analysis of this data produced the same structure as the best-fit structure at 456~K (Fig.~\ref{iv} and Table~\ref{params}). Second, we optimized the bulk-Ru lattice constant in the R-factor analysis of the 456~K data.  The minimum $R_p$ occurred at the room-temperature Ru lattice constant (within $\pm$0.2\%) and, again, the error bars were nearly the same as the analysis without this additional parameter.  Thus, we conclude that the first interlayer spacing of Ru is contracted by 3.5\% both at room temperature and at 456~K.

We next discuss the advantages of determining a surface structure from a single terrace instead of using multiple terraces.  First, for single-terrace data a larger energy range, $\Delta E$, can be obtained since symmetry-inequivalent beams are not averaged together. This larger energy range improves the reliability of the fit. Indeed, our energy range is larger than that used in previous LEED analysis\cite{FEIBELMAN1994}. To evaluate this potential error source, we performed an independent R-factor analysis of our experimental data where the experimental (10) and (01) beams were averaged. Compared to the single-terrace data set, the best fit for this symmetrized data set shows a slightly smaller interplanar contraction but a significantly larger error bar, $\delta d_1=-3.3(1.4)$\%. With this larger error, this result would be compatible with the previous experimental results.

An additional uncertainty in multiple-terrace data is that the two terrace types are usually taken as being present in equal abundance.  But this is not always the case.  Ru(0001) has a marked tendency to form ``double-steps,'' that is, one terrace type is considerably wider than the other terrace type. The narrow terrace type is bounded by two closely spaced ("double") steps.  Which terrace type is more abundant depends upon the local orientation of the steps.  Even for an averaging technique, such as obtained from a traditional diffractometer, the two terraces types may not be in equal abundance, giving an extra parameter to fit and increasing the error bars. This effect might explain the experimental observation of three-fold patterns on Ru surfaces by X-ray photoelectron diffraction\cite{Ruebush0a}. In fact, the change in symmetry has been employed to monitor the presence of double steps\cite{Menzel1995}. Double-steps may also introduce a non-negligible correction to the usual assumption of incoherent beam mixing in the theoretical 
I-V simulations, further reducing the reliability of the R-factor analysis.

\begin{table}

\begin{center}
\begin{tabular}{cc|cc} \hline \hline
 \multicolumn{2}{c|}{Structural Parameters [\AA]}&
               \multicolumn{2}{c}{Non-structural Parameters} \\ \hline
d$_{1}$  & 2.065 $\pm$ 0.02 & r$_m$        & 2.00 bohr  \\
d$_{2}$  & 2.140 $\pm$ 0.025& V$_i$        & -2.6 E$^{1/3}$ [eV] \\
d$_{bulk}$& 2.141            & $\Theta_D^1$ & 350 K \\
          &                  & $\Theta_D^2$ & 400 K \\
          &                  & $\Theta_D^{bulk}$ & 600 K \\ \hline
\end{tabular}
\end{center}
\caption{Optimized structural and non-structural parameters deduced in
this work. See text for further explanations.}
\label{params}
\end{table}

The final point concerns the role of adsorbed hydrogen in the experimental structures\footnote{The LEED analysis of Menzel {\it et al.} as well as the SXRD analysis of Baddorf {\it et al.} concluded that H adsorption was not responsible for the experimentally determined Ru(0001) first interlayer spacing\cite{Menzel1994,Baddorf2002}. However, the calculations of Feibelman {\it et al.}\cite{FEIBELMAN1994} show that H adsorption significantly decreases the surface contraction.}. We note that our structure, which was obtained from a crystal above room temperature, agrees extremely well with {\it ab-initio} calculations. Since there is no disagreement, we find no need to invoke impurity effects. To our knowledge, this is the only experimental result which corroborates the 3.5\% contraction deduced from the hydrogen-free DFT calculations.

\section{Summary}
Using a LEEM instrument we have performed, for the first time, a LEED I-V analysis from a 
single atomic Ru(0001) terrace. The experimental I-V curves were excellently fit by full dynamical calculations.  In the best-fit structure, the topmost Ru layer is relaxed inward by about 3.5\%, in excellent agreement with {\it ab-initio} calculations.  We suggest that determining structures from small areas has clear advantages even for surfaces that consist of a single structure (phase), such as Ru.

\begin{ack}
The authors thank Jose Emilio Prieto for useful
discussions. This research was partly supported by the U. S. Department of Energy, Office of
Basic Energy Sciences, Division of Materials Sciences under contract No. DE-AC04-94AL85000; by the Comunidad Aut\'{o}noma de Madrid through
Projects No.~GR/\-MAT/0155/2004 and GR/\-MAT/0440/2004; and by
the Spanish Ministry of Science and Technology through Projects
No.~MAT2003-08627-C02-02 and MAT2004-04348. F.E. gratefully acknowledges support from 
a fellowship from the Education and Science Spanish Ministry.
\end{ack}


\end{document}